\documentclass[a4paper,14pt,psfig,epsfig,amsmath,amssymb]{article}

\usepackage{amssymb,epsfig}
\usepackage{amsfonts}
\usepackage{amsmath}
\usepackage{graphics}

\begin{document}
\vspace{-15cm}
\title{Bimodality Phenomenon in Finite and Infinite Systems Within an Exactly Solvable Statistical Model}

\author{
 V.V.~Sagun$^1$, A.I. Ivanytskyi$^1$, K.A.~Bugaev$^{1,2}$, D.R. Oliinychenko$^{1,2}$ \vspace*{0.5cm}\\
$^1$Bogolyubov Institute for Theoretical Physics, National Academy \\ of Sciences of Ukraine, Kiev, Ukraine\\
$^2$Frankfurt Institute for Advanced Studies (FIAS), Goethe-University,\\ Ruth-Moufang Str. 1, 60438
Frankfurt upon Main, Germany
}



\maketitle

\begin{abstract}
 We present a few explicit counterexamples to the widely spread belief about
 an exclusive role of the bimodal nuclear fragment size distributions as the
 first order phase transition signal.  In  thermodynamic limit the bimodality may
appear at the supercritical temperatures due to the negative values of the surface tension coefficient. Such a  result is found within a novel exactly solvable formulation of the simplified statistical multifragmentation model based on  the virial expansion for a system of the nuclear fragments of all sizes. The developed statistical model corresponds to the compressible nuclear liquid with the tricritical endpoint located at one third of the normal nuclear density.
 Its exact solution for finite volumes demonstrates  the bimodal fragment size
distribution right inside  the finite volume analog of a gaseous phase.  These
counterexamples clearly demonstrate the pitfalls of  Hill approach to phase
transitions in finite systems.
\end{abstract}

\section{Introduction}

Despite many efforts the phase transition (PT) thermodynamics of finite systems is far from being completed.
Its consistent  formulation  remains a real theoretical challenge  for the researchers  working in statistical mechanics. 
On the other hand, nowadays it is of great practical importance since at  intermediate
and high energies  the modern nuclear physics  is dealing with the phase transformations of liquid-gas type  occurring in finite or even small systems. 
The central issue of this field  is related to a rigorous definition  of finite volume analogs of phases.

The first attempt  \cite{THill:1} to rigorously define the gaseous and liquid phases in finite systems was based on 
the properties of phases existing  in  infinite systems  in which two phases coexist at phase equilibrium 
 and generate two  local maxima, i.e. a bimodality,  of some order parameter. 
 Each maximum  is associated with a pure phase  
 \cite{THill:1}. 
 Since a few years ago such a concept of nuclear liquid-gas PT \cite{Bmodal:Chomaz03, Bmodal:Gulm07} completely dominates in nuclear physics of intermediate energies.
It considers  the bimodal distributions as a robust signal of a PT  in finite systems.
 However,  this concept does not seem to be correct since in a finite system an analog of mixed phase  is not just a simple mixture of two pure phases as it is explicitly shown within an exactly solvable statistical model  \cite{Bugaev:CSMM05, Sagun:Ivanytskyi13, Sagun:Bugaev13}.
The aim of this work  is to demonstrate that   in finite and infinite systems  the bimodal distributions can appear without a PT  and, hence,   they cannot serve as robust signal of a PT in finite systems.

\subsection{Constrained  SMM with the compressible nuclear matter}

The simplified statistical multifragmentation model (SMM) which has 
no Coulomb and  no asymmetry energy
was exactly solved in thermodynamic limit  in  \cite{Sagun:Bugaev00}, while 
its generalization constrained for  finite systems, the CSMM,  was solved   in \cite{Bugaev:CSMM05}.  
For a  volume  $V$ the grand canonical partition of the CSMM 
can be identically written as \cite{Bugaev:CSMM05, Sagun:Ivanytskyi13, Sagun:Bugaev13}
 \begin{equation}\label{SagunI}
{\cal Z}(V,T,\mu)~ = \sum_{\{\lambda _n\}}
e^{\textstyle  \lambda _n\, V }
{\textstyle \left[1 - \frac{\partial {\cal F}(V,\lambda _n)}{\partial \lambda _n} \right]^{-1} } \,,
\end{equation}
where  the set of  $\lambda_n$ $(n=0,1,2, 3,..)$ are all the roots of  the equation $\lambda _n~ = ~{\cal F}(V,\lambda _n)$.

The volume spectrum of our model ${\cal F}(V,\lambda)$ depends on the eigen volume $b = 1/ \rho_0 $ of a  nucleon at 
the normal nuclear density
$\rho_0\simeq 0.17$ fm$^3$ taken  at $T=0$ and zero pressure, mass $m \simeq 940$ MeV, degeneracy factor  $z_1 = 4$ of nucleons and it is defined as 
\begin{eqnarray}\label{SagunII}
&&\hspace*{-0.6cm}{\cal F}(V,\lambda)=\left(\frac{m T}{2 \pi}\right)^{\frac{3}{2}}z_1\exp \left\{\frac{\mu-\lambda T b}{T}\right\}+\hspace*{-0.1cm} \sum_{k=2}^{K(V)}\phi_k (T) \exp \left\{\frac{( p_L(T,\mu)- \lambda T)b k }{T} \right\}\,.~~~~~
\end{eqnarray}
Here $\phi_{k>1}(T)\equiv\left(\frac{m T }{2 \pi}\right)^{\frac{3}{2} } k^{-\tau}\,\exp \left[ -\frac{\sigma (T)~ k^{\varsigma}}{T}\right]$ is a reduced distribution function of the $k$-nucleon fragment, $\tau$ is the Fisher topological exponent and $\sigma (T)$ is the $T$-dependent surface tension coefficient.  Usually, the constant, parameterizing the dimension of surface in terms of the volume is  $\varsigma = \frac{2}{3}$. In the expression for ${\cal F}(V,\lambda)$    the  maximal size of  fragment is denoted as $K(V)$.
In the usual SMM \cite{Sagun:Bondorf:95} and  in its simplified version SMM the nuclear liquid pressure  $p_L^{SMM}=\frac{\mu+W(T)}{b}$ corresponds to an  incompressible matter.  Since  this is in contradiction with  the experimental  heavy ions collisions data \cite{Khan:2009}, 
here we analyze  the following equation of state with non-zero compressibility
\begin{equation}
\label{SagunIV}
p_L=\frac{ W(T) +  \mu + a_\nu ( \mu -\mu_0)^{\nu}}{b} \,
\end{equation}
which contains an additional term to the usual SMM liquid pressure.
Here an integer power is $\nu=2$ or $\nu=4$, $ W(T) = W_0 + \frac{T^2}{W_0}$ denotes  the usual  temperature dependent  binding energy per nucleon with $W_0 =  16$ MeV \cite{Sagun:Bugaev00}, while   the constants  $\mu_0 = - W_0$, $a_2 \simeq 1.233 \cdot 10^{-2}$ MeV$^{-1}$ and $a_4 \simeq 4.099 \cdot 10^{-7}$ MeV$^{-3}$  are fixed in order   to reproduce  the properties  of normal nuclear matter, i.e. at vanishing temperature  $T=0$ and normal nuclear density $\rho = \rho_0$ the liquid pressure must be zero. 
Under a new ansatz for  $p_L$  the nuclear liquid of CSMM becomes compressible \cite{Sagun:Bugaev13,Sagun:Sagun13}. 
A  careful analysis of the proposed parameterization \cite{Sagun:Sagun13} shows that it is fully consistent with the L. van Hove axioms of statistical mechanics \cite{Hove}. 

In addition to a  more general  parameterization of the bulk  free energy of nuclear  fragments we  also consider a more general parameterization of the surface tension coefficient 
\begin{equation}\label{SagunV}
 \sigma (T) =  \sigma_0 \left| \frac{T_{cep} - T }{T_{cep}} \right|^\zeta  {\rm sign} ( T_{cep} - T) ~,
\end{equation}
with  $\zeta = const \ge 1$, $T_{cep} =18$ MeV and $\sigma_0 = 18$ MeV the SMM. In contrast to the Fisher droplet model \cite{Fisher:67} and the usual SMM \cite{Sagun:Bondorf:95}, the CSMM surface tension (4) is negative above the critical temperature $T_{cep}$. An extended discussion on the validity of such a parameterization can be found in  Refs. \cite{Sagun:Ivanytskyi13,Sagun:Bugaev13}.

\subsection{Infinite system}

In the thermodynamic limit, i.e. for $V \rightarrow \infty$ and  $K(V) \rightarrow \infty$,
in the CSMM 
there is always a single solution $\lambda_0$ of the equation $\lambda _n~ = ~{\cal F}(V \rightarrow \infty,\lambda _n)$, but it can be of two kinds \cite{Bugaev:CSMM05}: 
 either the gaseous pole $ \lambda_0 (T, \mu) = p_g (T, \mu) / T$ for  ${\cal F}(V\rightarrow \infty,\lambda_0 - 0) < \infty$ or the liquid essential singularity $\lambda_0 (T, \mu)= p_L (T, \mu) /T $ for  ${\cal F}(V\rightarrow \infty,\lambda_0 - 0) \rightarrow  \infty$. 

\begin{figure}[!]
\begin{minipage}[h]{0.49\linewidth}
\center{\includegraphics[width=1.1\linewidth]{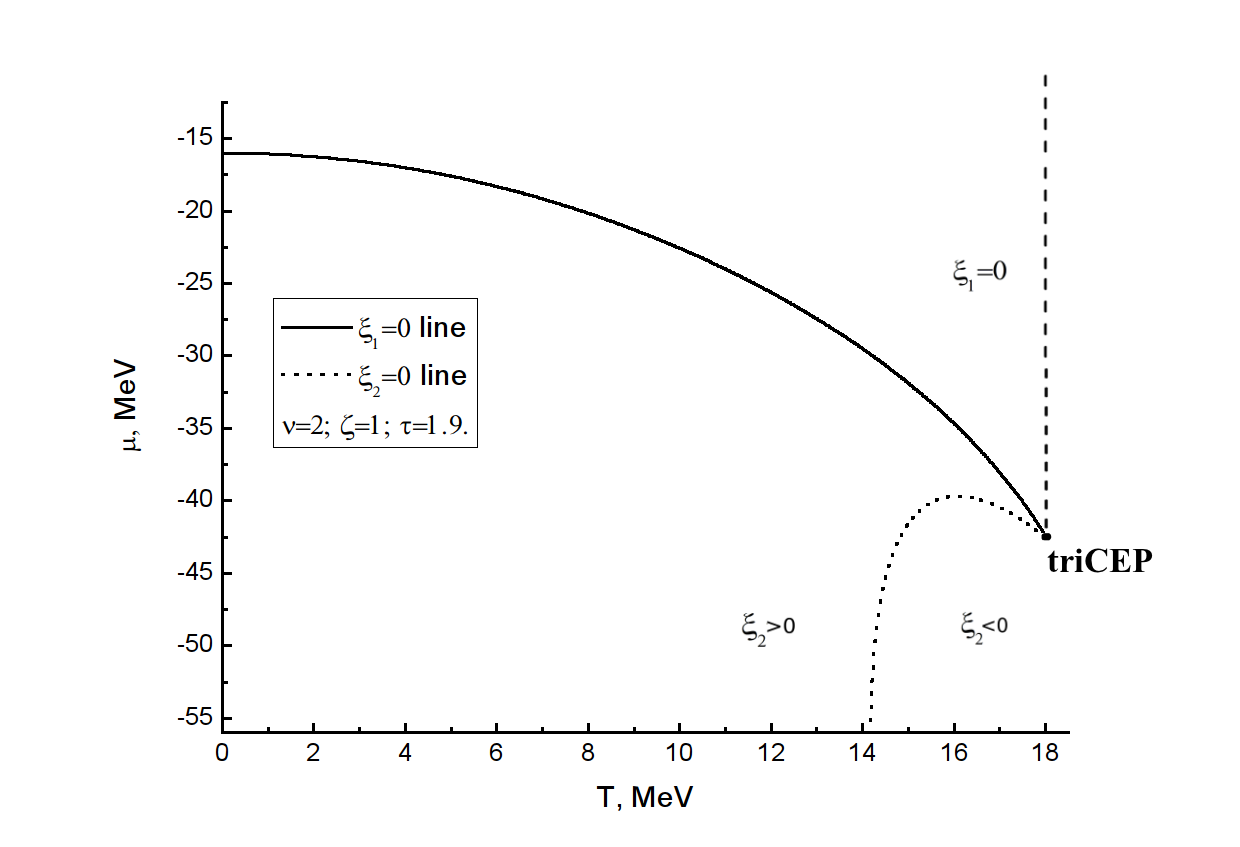}}
\end{minipage}
\hfill
\begin{minipage}[h]{0.49\linewidth}
\center{\includegraphics[width=1.1\linewidth]{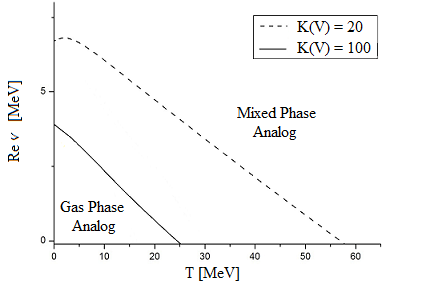}}
\end{minipage}
\caption{{\bf Left panel:}
Phase diagram in $T-\mu$ plane for the case $\nu=2$, $\tau=1.9$ with tricritical point at temperature $T_{cep}=18$ MeV in thermodynamic limit. The solid line corresponds to the 1-st order PT, the dashed line shows the 2-nd order PT, while at the dotted line the surface  tension coefficient vanishes. {\bf Right panel:} the finite volume analog of the phase diagram in $T-Re(\nu_{1})$ plane for given values of $K(V)=20$ (dashed curve) and $K(V)=100$ (solid curve). Below each of these phase boundaries there exists a gaseous phase only, but at and above each curve there are three or more solutions of the system (\ref{SagunVI}, \ref{SagunVII}).}
\label{Fig:sagunI}
\end{figure}

This  model has a PT which  occurs when the gaseous pole is changed by  the liquid essential singularity or vice versa. The PT curve  $\mu = \mu_c (T)$ is a solution of  the equation $p_g (T, \mu) = p_L (T, \mu) $, which is just the Gibbs criterion of  phase equilibrium. The properties of a PT are defined only by the liquid phase pressure  $p_L(T, \mu)$ and   by the temperature dependence of  surface tension $\sigma(T)$. The   phase diagram of the present model in thermodynamic limit in the plane of baryonic chemical potential $\mu$ and temperature $T$ is shown in the left panel of  Fig.~\ref{Fig:sagunI}.

\subsection{Finite system}

The treatment of the model for  finite volumes is more complicated, since the roots  $\lambda _n$ of   (\ref{SagunI})  have not only the real part $R_n$, but an imaginary  part $I_n$ as well ($\lambda _n = R_n + i I_n$). Therefore, 
equation for $\lambda_n$ can be cast as a system of coupled transcendental equations for $R_n$ and $I_n$ 
\begin{eqnarray}\label{SagunVI}
&&\hspace*{-0.cm} R_n = ~ \sum\limits_{k=1}^{K(V) } \phi_k (T)
~ \exp \left[ \frac{Re( \nu_n)\,k}{T} \right]  \cos(I_n b k)\,,
\\
\label{SagunVII}
&&\hspace*{-0.cm} I_n = - \sum\limits_{k=1}^{K(V) } \phi_k (T)
~\exp \left[ \frac{Re( \nu_n)\,k}{T} \right]  \sin(I_n b k)\,,
\end{eqnarray}
where  for convenience we introduced   the following  set of  the effective chemical potentials  $\nu_n $ 
\begin{equation}\label{SagunVIII}
\nu_n  \equiv  \nu(\lambda_n ) = p_l(T,\mu) b  - (R_n + i I_n) b\,T  \,,
\end{equation}
and the reduced distribution for nucleons $\phi_1 (T) = \left(\frac{m T }{2 \pi} \right)^{\frac{3}{2} }  z_1 \exp((\mu -  p_l(T,\mu) b)/T)$.

Consider the real root $(R_0 > 0, I_0 = 0)$, first. 
The real root $\lambda_0 = R_0$  of  the CSMM exists for any $T$ and $\mu$.
From (\ref{SagunI}) and  (\ref{SagunVI}) for $R_n = R_0$ and $I_0=0$ one can see that  $T R_0$ is  a constrained grand canonical pressure of the mixture of ideal gases with the chemical potential $\nu_0$. 
Hence, a single real solution $\lambda_0 = R_0$ with $I_0=0$ of the system (\ref{SagunVI}, \ref{SagunVII})
corresponds to a gaseous phase (for more details see \cite{Sagun:Bugaev00}). 
 If for some thermodynamic parameters we have a  real solution $\lambda_0$ and any finite  number  $ n=1, 2, 3, ...$ of  the  complex conjugate pairs of roots $\lambda_{n\ge 1}$,  then such a system  corresponds to a finite volume analog of mixed phase \cite{Sagun:Bugaev00}. 
 Note that, each pair of  complex conjugate roots $\lambda_{n\ge 1}$ represents  a 
 metastable state   with a complex value of  chemical potential $\nu_n$. 
 Since $\nu_{n1} \neq  \nu_{n2 \neq n1}$ these metastable states are
 not in a true chemical equilibrium with the gas and with each other. 
 A finite system analog of a  liquid  phase  corresponds to an infinite number of the complex roots of the system  
 (\ref{SagunVI}, \ref{SagunVII}), but  in finite system it exists at infinite pressure only.  Using these definitions, one can build up the finite system analog of the $T-\mu$ phase diagram (see the right panel of  Fig.1).
 
Therefore, in contrast to assumptions of Refs. \cite{Bmodal:Chomaz03, Bmodal:Gulm07}, in finite systems the pure liquid phase cannot exist at finite pressures. Instead, in finite system and finite pressures we are dealing 
with  the  finite volume analogs of gaseous or  mixed phases   \cite{Bugaev:CSMM05}.


\subsection{Bimodality phenomenon  in  finite and infinite systems}

In this section we  discuss  another  typical mistake of the approaches \cite{Bmodal:Chomaz03, Bmodal:Gulm07} based on the bimodal  properties of  the first order PT in finite systems. 
 The authors of  \cite{Bmodal:Chomaz03, Bmodal:Gulm07}  implicitly assume  that, like in the infinite systems, in finite systems there exist exactly  two `pure' phases  and  they exactly  correspond to  two peaks in the bimodal distribution of the order parameter.
As two counterexamples to these assumptions we present the  bimodal fragment distributions obtained for an infinite system at the supercritical temperature where the surface tension coefficient is negative (the left panel of  Fig.2) and the one obtained
inside the  finite volume analog of  a gaseous phase corresponding to 
positive values of the effective chemical potential  $\nu_0$ (the right panel of  Fig.2).
As one can see from Fig.2, in contrast to expectations of  \cite{Bmodal:Chomaz03, Bmodal:Gulm07}, the  bimodal fragment distributions occur  without a  PT. 


\begin{figure}[!]
\begin{minipage}[h]{0.49\linewidth}
\center{\includegraphics[width=1.12\linewidth]{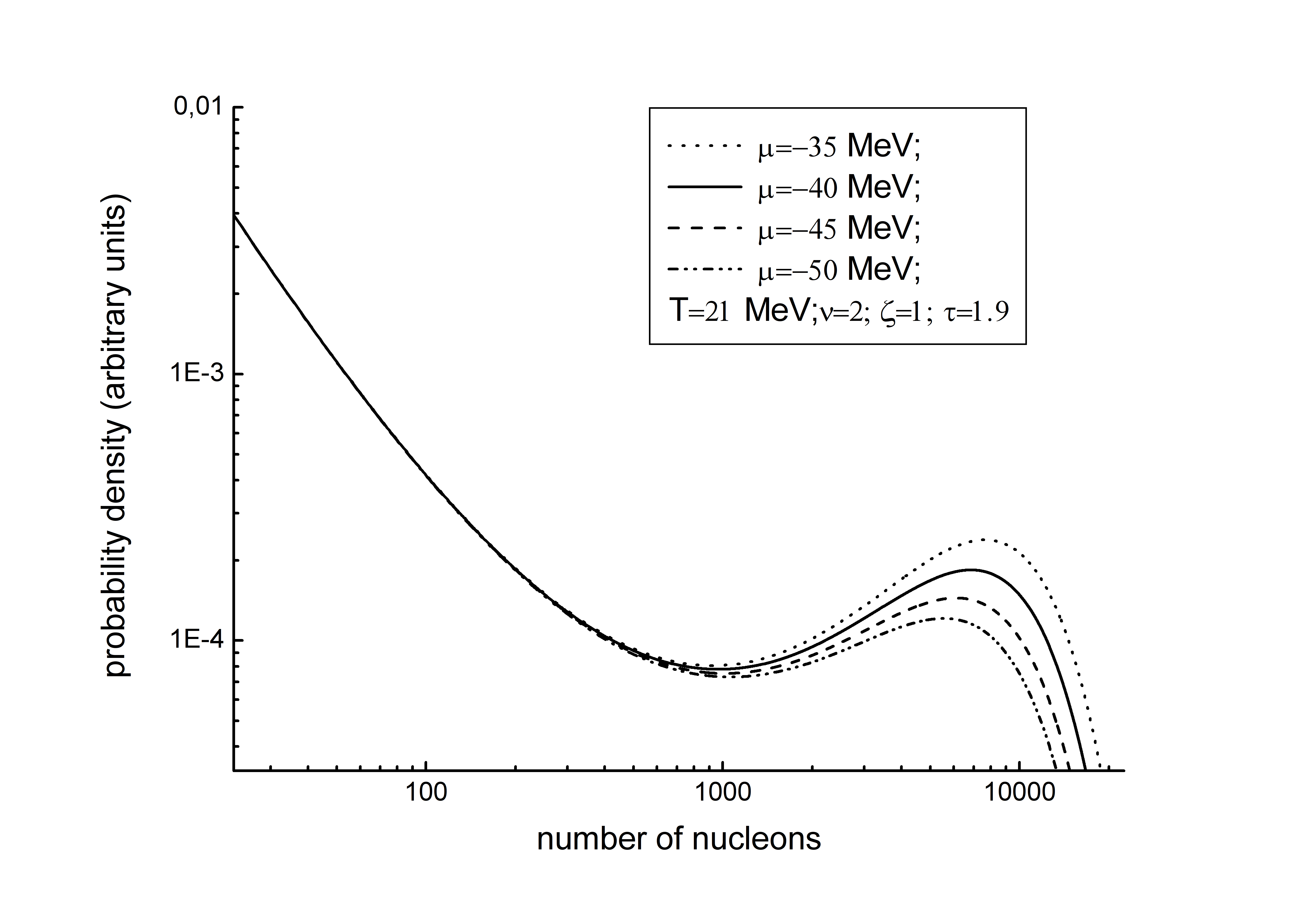}}
\end{minipage}
\hfill
\begin{minipage}[h]{0.49\linewidth}
\center{\includegraphics[width=1.0\linewidth]{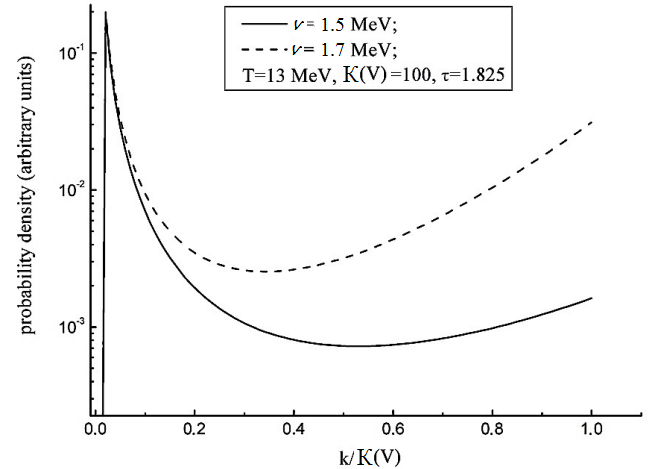} }
\end{minipage}
\caption{{\bf Left panel:}
 Fragment size distributions of the model are  shown for a fixed temperature $T=21$ MeV and four values of the   baryonic chemical potential $\mu$  for an infinite system. This region of phase diagram   is characterised by the negative surface tension coefficient which prevents an existing of a  PT.
{\bf Right panel:} 
Bimodal distributions existing inside the finite system analog of gaseous phase for a fixed temperature $T=13$ MeV and different values of the effective chemical potential $\nu_{0}$. Even in the region of fragments gas we observe a bimodal like shape of the fragment distribution. The maximal size of nuclear fragment is $K(V)$=$k$=100 nucleons.}
\label{Fig:sagunII}
\end{figure}

\section{Conclusions}

A novel  version of the CSMM is presented here. Its detailed analysis 
is performed in order to clarify an origin
of the bimodality appearing  both in finite and in infinite systems. 
An exact analytical solution of the present model allows  us to perform a robust analysis
of the fragment size distributions in the regions where there is and there is no  PT. 
It is shown that the fragment size 
distribution can be bimodal-like  inside of   the finite volume  analog of gaseous phase. Also we
demonstrate  that a bimodal fragment size 
distribution  can be  caused by negative values of the surface tension and, hence, 
it is not a robust signal  of  PT existence  in finite systems.

\mbox{}\\
\hfill\\
\noindent
{\bf Acknowledgments}
The authors appreciate the valuable comments of  I. N. Mishustin  and  G. M. Zinovjev. 
The authors acknowledge  a  support of the Program `On Perspective Fundamental Research in High Energy and Nuclear Physics' launched by the Section of Nuclear Physics  of NAS of Ukraine and a support provided  by the State Program of Ukraine  on the GRID technology implementation.
K.A.B. and D.R.O.    acknowledge  a partial support provided by the Helmholtz 
International Center for FAIR within the framework of the LOEWE 
program launched by the State of Hesse.

\begin{footnotesize}

\end{footnotesize}

\end{document}